\documentclass[12pt,fleqn]{article}
\pdfoutput=1

\usepackage{eurosym}
\usepackage{amsmath,amsfonts,amssymb,amsthm}
\usepackage{latexsym}
\usepackage{mdwlist}
\usepackage{a4wide}
\usepackage{ae,aecompl}
\usepackage{titlesec}
\usepackage{graphicx}
\usepackage{epstopdf}
\usepackage[singlelinecheck=false]{caption}
\usepackage[round]{natbib}
\usepackage{url}
\usepackage{pdfpages}
\usepackage{subfig}
\usepackage[breaklinks=true]{hyperref}
\usepackage{setspace}
\usepackage{xcolor}
\usepackage{soul}
\usepackage[anythingbreaks]{breakurl}
\usepackage{array}
\usepackage{multirow}

\makeatletter

\newtheorem{proposition}{Proposition}

\newtheorem{example}{Example}

\newcommand{\Xomit}[1]{}

\titleformat*{\section}{\large\bfseries}
\titleformat*{\subsection}{\normalsize\bfseries}

\begin{document}

\title{OPTIMAL TRADE-OFF BETWEEN ECONOMIC ACTIVITY AND HEALTH DURING AN EPIDEMIC\footnote{We are grateful to J\"{o}rgen Weibull for valuable comments and suggestions and, in particular, for his detailed notes on the epidemiological SI-model considered in this paper.}}
\bigskip
\author{Tommy Andersson\footnote{Lund University and Stockholm School of Economics, Departments of Economics. E-mail: \texttt{tommy.andersson@nek.lu.se}.}, Albin Erlanson\footnote{University of Essex, Department of Economics. E-mail: \texttt{albin.erlanson@essex.ac.uk}.}, Daniel Spiro\footnote{Uppsala University, Department of Economics. E-mail: \texttt{daniel.spiro@nek.uu.se}.} \space and Robert \"{O}stling\footnote{Stockholm School of Economics, Department of Economics. E-mail: \texttt{robert.ostling@hhs.se}.}}
\date{\today}

\maketitle

\begin{abstract}
\noindent This paper considers a simple model where a social planner can influence the spread-intensity of an infection wave, and, consequently, also the economic activity and population health, through a single parameter. Population health is assumed to only be negatively affected when the number of simultaneously infected exceeds health care capacity. The main finding is that if (i) the planner attaches a positive weight on economic activity and (ii) it is more harmful for the economy to be locked down for longer than shorter time periods, then the optimal policy is to (weakly) exceed health care capacity at some time.

\medskip


\end{abstract}

\section{Introduction}
A central part of many countries' policies to tackle the Covid-19 pandemic
has been to \textquotedblleft flatten the curve.\textquotedblright\ That is,
a more gradual uptick of infected persons prevent health care systems to be
overburdened and save human lives. For example, \citet{Greenstone} \citep[building on][]{Ferguson} estimate that 630,000 lives in the US could be saved by social distancing
policies assuring that intensive care units are not overwhelmed during the
peak of the Covid-19 pandemic. At the same time, slowing down disease transmission appear to cause large economic costs due to a fall in both consumption and production.
\citet{Fernandes} estimates the costs of the Covid-19 outbreak for 30
countries under different scenarios, and finds a median decline in GDP in
2020 of 2.8 percent, but that GDP can fall by more than 10--15 percent in
some scenarios. Many have therefore concluded that the Covid-19 outbreak
involves a key trade-off: a higher spread-intensity is advantageous for
economic activity, but disadvantageous for population health.

This paper analyzes the trade-off between reduced economic activity and population health in a simple and tractable model. In contrast to most previous work, we simplify the epidemiological model by only allowing two states: individuals are either susceptible or infected. This modelling choice implies that the whole population is eventually
infected and that there is no death or recovery from the infection.\footnote{In the SIR-model \citep{KermackMcKendrick}, the letters S, I and R stand for susceptible, infectious and recovered, respectively. The considered model is a SI-model since individuals never recover from the infection.\label{Footnote:SIR}} This is overly simplistic for studying disease transmission more generally, but we believe it can be an useful simplification when integrating epidemiological and economic models. In particular, our model allows the social planner to influence how quickly the infection spreads, and therefore also the economic activity and population health, by choosing a single parameter. A lower spread-intensity increases economic activity, but harms population health if the number of infected at the peak of the epidemic exceeds health care capacity.

We first show that if the social planner only puts weight on population health, health care capacity will never be exceeded, which is in line with arguments behind \textquotedblleft flattening the curve\textquotedblright\ policies. The same conclusion holds if the social planner is also concerned about upholding economic activity, but production is not affected by how quickly the disease is spreading. In more realistic scenarios where the social planner attaches a positive weight on economic activity and it is more harmful for the economy to be locked down for longer than shorter time periods (e.g., because social distancing policies are more harmful for the economy the longer time they are enforced), the optimal policy is to (weakly) exceed health care capacity during the some time of the epidemic.

Our model is deliberately kept stylized and abstracts from several relevant considerations. The trade-off between economic activity and population health would arise also in more elaborate models, but the finding that it is optimal to (weakly) exceed health care capacity is more sensitive to modelling assumptions. There are several possible reasons why a slower spread of the disease (a flatter curve) below the health care capacity constraint may be optimal in a richer model. For example, if patients recover from the disease and develop immunity (as in a SIR-model), a slower spread of the desease may limit the share of the population that is eventually infected. A slower spread may also be optimal if population health is negatively affected when more individuals are
simultaneously infected also below the health care capacity constraint. Finally, the possibility that a vaccine or better medical treatments becomes available provides additional incentives to delay the epidemic.

Although our model implies a sharp trade-off between output and population health, there are mechanisms that could mitigate that trade-off. In the context of our model, it would be beneficial to develop policies and technologies that can lower spread intensity while harming production less, perhaps using testing \citep[e.g.][]{Berger}. Another potential mechanism is that high disease transmission may reduce economic activity because people spontaneously limit consumption and reduce labor supply in fear of being infected also in the absence of a policy response \citep{Fenichel,Atkeson,Eichenbaum}. Finally, because we do not explicitly incorporate mortality in our model, one apparent mechanism that dampens the trade-off is that high mortality reduces population size and thereby production. However, we believe this latter channel to be of minor importance in the context of Covid-19 due to relatively low mortality during productive years.

Our paper is a related to a number of recent papers, e.g., \citet{Alvarez}, \citet{Eichenbaum}, \citet{Gollier} and \citet{Jones}, that combine the canonical epidemiological SIR-model \citep{KermackMcKendrick} with macroeconomic models to analyze how policymakers should optimally respond to a pandemic while taking both economic activity and population health into account. For example, \citet{Eichenbaum} assumes that social distancing reduces consumption and labor supply, which limits spread of the
disease and reduce economic activity. Social distancing hence exacerbate the recession but raise welfare by reducing the number of pandemic-related deaths. \citet{Eichenbaum} solve the model numerically and calibrate it to the Covid-19 pandemic and find that it is optimal to introduce large-scale containment measures even if they result in a sharp (and sustained) output drop. Several other papers using similar modelling approaches calibrated to the Covid-19 pandemic also conclude that drastic front-loaded policies are optimal \citep{Alvarez,Gonzalez-Eiras,Jones}.

Our modelling approach is simpler which allows us to derive theoretical results without numerical calibration. In this respect, our model is more similar to earlier work in epidemiology starting with \citet{Abakuks} that study optimal disease control under resource constraints \citep[see][for a recent survey]{Nowzari}. Another related paper in the domain of purely analytical results is \cite{behncke2000optimal} who shows existence of optimizers for a number of different policies. More recently, \cite{Kruse} analyze optimal suppression when minimizing the total number of infected over time (i.e., a different objective function than in our model) subject to a cost of doing so, and \cite{MorrisEtAl} analyze how to minimize the peak of the infection curve when marginal cost of suppression is zero.

The paper closest to ours is \citet{Miclo} who focus on optimal suppression in order to not overwhelm health care capacity. They show that the optimal policy is time-varying. In contrast to them, the planner in our model is restricted to time-variant polices, but can exceed health care capacity at a cost. Among the recent numerically-oriented papers \cite{favero2020restarting} is the closest as they also take into account that harm increases substantially above health care capacity.

\section{The Model}
An infection spreads in the population and a social planner must decide on a policy regarding the spread-intensity. The planner takes both production and population health into consideration and faces a trade-off: a higher spread-intensity implies that the economy needs to be locked down for a shorter period of time, but it imposes a higher stress on the health care system. To model this trade-off, we first introduce a simple infection model where a social planner can influence the spread-intensity through a single parameter.

\subsection{The Infection Model}
At any time $t\geq 0$, $x(t) \in (0,1)$ represents the share of the population that have been infected before time $t$. Assume uniform pairwise random matching in the population, and that the infection spreads with probability $p$ when an infected individual meets a susceptible individual (even if the infection occurred a long time ago). Assume also that the time-rate of pairwise meetings is $m>0$. In this simple infection model, there are only two types of individuals: susceptible and infected. In particular, there is no death or recovery from the infection, i.e., the infection model is what epidemiologists refer to as a SI-model (see footnote \ref{Footnote:SIR}). A social planner can affect both $p$ and $m$, e.g., by different containment policies, social distancing rules and various hygiene advice campaigns, but only at time $t=0$ . The spread-intensity parameter is given by $a=pm$.

The mean-flow dynamic of the infection over time is then given by the following ordinary differential equation:\footnote{To the best of our knowledge the first  time the logistic model was used to describe a population growth, as the one we have above, was by \cite{Verhulst38} and further developed by the same author in \cite{Verhulst45}.}
   \begin{equation}
     \dot{x}=a x(1-x), \label{EQ:PDF}
   \end{equation}
\noindent with initial value $x(0)\in (0,1)$. Equation (\ref{EQ:PDF}) uniquely determines the dynamic evolution of the disease, and its solution is given by:
   \begin{equation}
     x(t) =\frac{e^{at}}{e^{ab}+e^{at}},\label{EQ:CDF}
   \end{equation}
\noindent where:\footnote{Note that $e^{ab}=\frac{1}{x(0)}-1$. This way of writing the constant in the ODE simplifies later arguments.}
   \begin{equation}
     b=\frac{1}{a}\ln \left(\frac{1}{x(0)}-1\right).\label{EQ:b}
   \end{equation}
\noindent From equation (\ref{EQ:CDF}), it follows that:
   \begin{equation*}
     \dot{x}(t)=\frac{ae^{at}(e^{ab}+e^{at})-ae^{at}e^{at}}{(e^{ab}+e^{at})^2}=ae^{ab}\frac{e^{at}}{(e^{ab}+e^{at})^{2}}.
   \end{equation*}
\noindent Note that $\dot{x}(t):\mathbb{R}\rightarrow \mathbb{R}_{+}$ is a probability density function that describes the infection wave (in this case, a hump-shaped function, see the dashed-dotted lines in Figure \ref{fig:example}), and its integral $x(t):\mathbb{R}\rightarrow [0,1]$ a cumulative density function. Consequently, for a given spread-intensity parameter $a$, the ``size'' of the infection wave at time $t$ is given by $\dot{x}(t)$, and the ``peak'' of the wave occurs at the time $\hat{t}$ where $\ddot{x}(\hat{t})=0$. It can be verified that $\hat{t}=b$ where $b$ is given by equation (\ref{EQ:b}) for any value of $a$. Because the value of $b$ is proportional to $1/a$, the greater $a$ is the smaller $\hat{t}=b$ is. A high spread-intensity rate therefore yields an early peak of the infectious disease.

\subsection{Maximizing Social Welfare During a Pandemic}
A social planner determines the optimal spread-intensity of the pandemic by taking both economic activities and population health into consideration. To spell out this trade-off formally we introduce a production function $y$  and a health function $h$  as a measure of how the economic activity and the health is affected by the pandemic, respectively.

\subsubsection{Production and Health}
We begin by specifying the production function $y(t\mid b)$. Similar to for example \citet{Eichenbaum}, we consider a problem in the short run (see also footnote \ref{Footnote:short_run}), so capital is fixed and we thus only need to consider labor when specifying the production function. In particular, it is assumed that production at time $t$ depends on the share of non-infected individuals (i.e., the available labor force at time $t$) together with a continuous and differentiable function $g:[\underline{b}, T]\rightarrow \mathbb{R}_+$. The idea is that $g$ controls how production is affected by the ``length'' of the period until the pandemic hits its peak at $b$.
 It is assumed that $g(\underline{b})\geq 1$ and $g^\prime(b)\geq 0$ for all $b\geq\underline{b}$. These assumptions on $g$ captures that the further away in time the peak of the pandemic is, the more harmful it is for production. The following reverse hump-shaped production function (see the dashed lines in Figure \ref{fig:example})  describes  this relation between the peak of the pandemic and the output produced  in the economy
   \begin{equation}\label{EQ:prod}
     y(t\mid b)=1-g(b)\dot{x}(t).
   \end{equation}
\noindent To ensure that $y(t\mid b)\geq 0$, it is also assumed that $g(b)\dot{x}(t)\leq 1$  for all $(b,t)\in [\underline{b},T]\times[0,T]$. From the production  function \eqref{EQ:prod} it follows that in the normal state of the economy, the entire population is working and produces an output equal to 1 (this is only a normalization, any positive number instead of 1 is fine). When the share of the population that has been infected approaches 1, the production again approaches the normalized output of 1.\footnote{The results presented in the next section will not qualitatively change if we add the assumption that immunity is reached when a given proportion of the population has been infected, and that the production instead equals 1 as soon as immunity is reached.} In all other time periods $t$, the production level depends both on the share of infected individuals  $\dot{x}(t)$ and the function $g(b)$ as specified in equation \eqref{EQ:prod}.

Let us now look at the  health function $h(t\mid b)$ that determines the impact on health from the pandemic. We assume that there is a fixed capacity in the health care system, denoted by $c\in[0,1]$, so if the peak is ``too high,'' not all infected individuals can get proper health care at all times $t$.  In fact, it can be shown that if $a>4c$, then the health care capacity is (weakly) exceeded in the time interval $[t_l,t_r]$ where (see also the left panel of Figure \ref{fig:example}):
   \begin{eqnarray}
     t_l&=&b+\frac{1}{a}\ln\left(-\frac{2c-a}{2c}-\sqrt{\left(\frac{2c-a}{2c}\right)^2-1}\right),\label{EQ:left}\\
     t_r&=&b+\frac{1}{a}\ln\left(-\frac{2c-a}{2c}+\sqrt{\left(\frac{2c-a}{2c}\right)^2-1}\right)\label{EQ:right}.
   \end{eqnarray}
\noindent From these conditions and equation (\ref{EQ:b}), it follows that the ``height'' of the peak is exactly equal to the  health care capacity at time $\hat{t}=b^\ast$, i.e., $\dot{x}(b^\ast)=c$, when:
\begin{equation*}
b^\ast=\frac{1}{4c}\ln\left(\frac{1}{x(0)}-1\right).
\end{equation*}
\noindent For $a<4c$, the capacity constraint $c$ is never binding and all infected patients can receive treatment.

Infected individuals need medical treatment at the time $t$ when they are infected but not before or after. From the social planner's perspective, this means that the health measure $h(t\mid b)$ in period $t$ is given by the share of the population that has not yet been infected, or has been infected before time $t$, or are infected at time $t$ but receives proper health care:
   \begin{equation}\label{EQ:health}
      h(t\mid b)=\left\{\begin{tabular}{ll} 1 & if $t\in [0, t_l]$ or $t\in [t_r,T]$,\\ $1-(\dot{x}(t)-c)$ & if $t\in (t_l, t_r)$. \end{tabular}\right.
   \end{equation}
\noindent  Note also that $t_l$ and $t_r$ are functions of $b$.

\subsubsection{The Social Welfare Function}
Having specified how the economy and the health in the society is affected by the pandemic we can now write down the welfare in the society at time $t$ as:
   \begin{equation}
     w(t\mid b)=\lambda y(t\mid b)+(1-\lambda) h(t\mid b),\label{EQ:welfare_easy}
   \end{equation}
\noindent where $y(t\mid b)$ is the production function, $h(t\mid b)$ is the health function, and $\lambda\in[0,1]$ is a welfare weight reflecting the importance attached to production and health. The total welfare during the pandemic is obtained by integrating the welfare measure from time 0  up to some given time $T$,\footnote{It is not immediately clear how to choose $T$. In the remaining part of the paper, it is assumed that $T$ is a ``sufficiently large'' constant. This is one of three natural choices of $T$ listed by \citet[][p. 428]{HansenDay}. The other two are (i) $T\rightarrow \infty$ and (ii) some constant $I_{\min}$ indicating the first time period when the number of infected individuals is sufficiently large to end the pandemic. All results presented in this paper holds qualitatively also for the two alternative definitions of $T$. Note also that because we consider a short-run problem, there is no need to introduce discount factors.\label{Footnote:short_run}}
\begin{equation}\label{EQ:Welfare_function}
W(b)= \lambda\int_{0}^{T}y(t\mid b)dt+(1-\lambda)\int_{0}^{T}h(t\mid b)dt.
\end{equation}
The (short-run) objective for the planner is to select a spread-intensity that maximizes welfare. For convenience, we shall describe the planner's problem in terms of deciding on the time $\hat{t}$ where the infection wave peaks. Note also that one can equivalently consider the problem of selecting the optimal spread-intensity parameter $a$ since the exact relationship between $a$ and $b$ is given by equation (\ref{EQ:b}). As it is likely that it is practically difficult for the social planner to spread the disease ``very fast,'' we shall assume that the peak cannot occur before some point in time $\underline{b}>0$. We are, however, agnostic about how close in time $\underline{b}$ is to $t=0$.

The planner's objective to maximize the social welfare function (\ref{EQ:Welfare_function}) can be written as\footnote{In case the capacity not is exceeded for any $t\geq 0$, i.e., when there is no solution to equations (\ref{EQ:left})--(\ref{EQ:right}), this expression can be simplified. See equation (\ref{EQ:not_exceeded}) in Section \ref{SEC:proofs}.}
\begin{equation*}
     \max_{b\in[\underline{b},T]} W(b)=\lambda\int_{0}^{T}(1-g(b)\dot{x}(t))dt+(1-\lambda)\left(\int_{0}^{t_l}1dt+\int_{t_l}^{t_r}(1-(\dot{x}(t)-c))dt+\int_{t_r}^{T}1dt\right).
\end{equation*}
\noindent Thus, for a given welfare weight $\lambda\in[0,1]$, the objective for the social planner is to decide on the time where the infection wave peaks to maximize the social welfare given by (\ref{EQ:Welfare_function}).

\section{Results}
 The first result concerns the two extreme cases where the social planner puts all weight on either production or health.
   \begin{proposition}\rm\label{PROP:extreme_cases}
     Suppose that $b\geq \underline{b}$ and $g^\prime(b)>0$ for all $b\geq \underline{b}$. If (i) $\lambda=0$, then any $b\geq b^\ast$ maximizes the welfare function (\ref{EQ:Welfare_function}), and if (ii) $\lambda=1$, then $b=\underline{b}$ maximizes the welfare function (\ref{EQ:Welfare_function}).
   \end{proposition}

\noindent The first part of the proposition states that if the social planner only is concerned about health, the optimal policy is to never exceed the health care capacity at any time. The second part of the proposition states that if the social planner only is concerned about production, the optimal policy is to make the infection wave peak as soon as it is feasible.

\begin{figure}[h!]
    \centering
    {{\includegraphics[width=7.5cm]{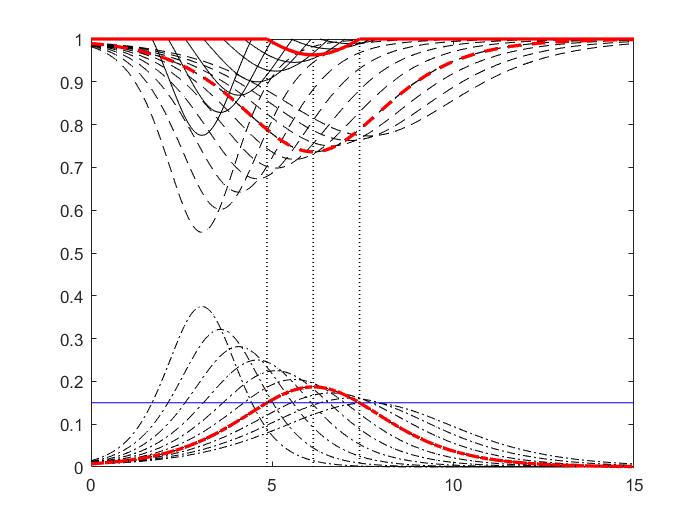}}}
    \qquad
    {{\includegraphics[width=7.5cm]{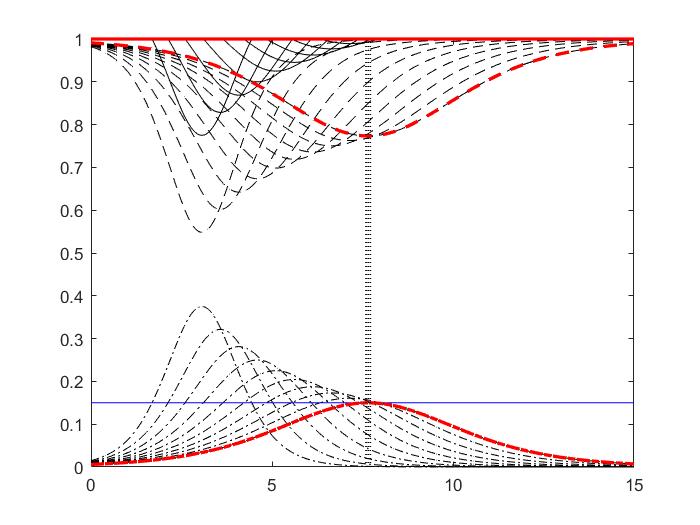}}}
    \caption{\small{In the left panel, the health care capacity is exceeded at the welfare maximizing solution ($b=6.14$). In the right panel, the ``height'' of the infection wave at the welfare maximizing solution ($b=7.66$) equals health care capacity (0.15).}}
    \label{fig:example}
\end{figure}

We next state another special case, namely the case when $g'(b)=0$ for all $b\geq \underline{b}$, i.e., when production is equally affected independently of the spread-intensity and when in time the pandemic peaks. In this case, the optimal policy is again to never exceed the health care capacity at any time.
   \begin{proposition}\rm\label{PROP:g_1}
     Suppose that $\lambda\in[0,1]$, and that $g'(b)=0$ for all $b\geq \underline{b}$. Then $b\geq b^\ast$ maximizes the welfare function (\ref{EQ:Welfare_function}).
   \end{proposition}
\noindent The above two propositions states that if the social planner only values health or if the economy is equally affected independently of when the infection wave peaks, the optimal policy is to never exceed health care capacity. However, the assumption that $g'(b)=0$ is rather unrealistic since this means that it is not more harmful for the economy to be locked down for longer than shorter time periods and that the social planner cannot affect the function $g$ by any policy measures. If these assumptions are dropped and if, in addition, the planner attaches a positive weight on production, Proposition \ref{TH:main} and Example \ref{Exemple} show that the optimal policy is to (weakly) exceed health care capacity.
   \begin{proposition}\rm\label{TH:main}
      Let $\lambda \in(0,1)$, and suppose that $g^\prime(b)>0$ for all $b\geq \underline{b}$. If $b$ maximizes the social welfare function (\ref{EQ:Welfare_function}), then it cannot be the case that $b>b^\ast$.
   \end{proposition}
   \begin{example}\label{Exemple}\rm
      Suppose that $x(0)=0.01$, $T=15$, $\underline{b}=3.06$, and $c=0.15$.\footnote{Note that the results in the example will not change if $\underline{b}<3.06$. The only thing that will change in Figure \ref{fig:example} is that additional curves has to be added to the left of $t=3.06$ but they will not be a solution to the planner's optimization problem for the given parameter values.} If $g(b)=1+\frac{b}{T}$ for all $b\geq \underline{b}$ and $\lambda=0.5$, the welfare maximizing peak of the infection wave occurs at $b=6.14$ implying that the health care capacity is exceeded in the interval $[4.86,7.42]$. This is illustrated in the left panel of Figure \ref{fig:example} where the infection waves (dashed-dotted lines) are illustrated in the bottom of the figure, and the production functions (dashed lines) and the health functions (solid lines) are illustrated in the top of the figure for 11 different values of $b$ between 3.06 and 7.66. The corresponding functions for the optimal $b=6.14$ are marked in red color.

      The right panel of Figure \ref{fig:example}, illustrates the situation for $\lambda=0.05$. In this case, the welfare maximizing policy is to set the peak of the infection wave at $b=7.66$, i.e., at the time where the ``height'' of the infection wave equals the health care capacity ($c=0.15$).\hfill $\square$
   \end{example}
\noindent Finally, we note that the if the social planner increases the welfare weight $\lambda$ or health care capacity $c$, the optimal value of $b$ weakly decreases. Thus, if the social planner attaches more weight on production or if health care capacity increases, the optimal policy is to select an infection peak closer in time.

\section{Proofs}\label{SEC:proofs}
Because all proofs are based on the same ideas, we start by stating some general remarks that will be useful in all of the proofs.

Note first that $t_r>b>t_l>\underline{b}$ for all $b\in (\underline{b},b^\ast)$ since capacity $c$ is exceeded for all $b\in [\underline{b},b^\ast)$, and $t_l=t_r=b>\underline{b}$ for $b=b^\ast$ since the ``height'' of the peak equals $c$ for $b=b^\ast$. Hence, the welfare function (\ref{EQ:Welfare_function}) is for any $b\in [\underline{b},b^\ast]$ given by:
   \begin{eqnarray}
     W(b)&=&\lambda\int_{0}^{T}y(t\mid b)dt+(1-\lambda)\int_{0}^{T}h(t \mid b)dt,\nonumber \\
         &=&\lambda\int_{0}^{T}(1-g(b)\dot{x}(t))dt+(1-\lambda)\left(\int_{0}^{t_l}1dt+\int_{t_l}^{t_r}(1-(\dot{x}(t)-c))dt+\int_{t_r}^{T}1dt\right),\nonumber\\
         &=&\lambda\left[t-g(b)x(t)\right]_0^T+(1-\lambda)\left(\left[t\right]_0^{t_l}+\left[t+ct-x(t)\right]_{t_l}^{t_r}+\left[t\right]_{t_r}^T\right),\nonumber\\
         &=& T - \lambda g(b)(x(T)-x(0)) +(1-\lambda)(c(t_r-t_l)-(x(t_r)-x(t_l))).\label{EQ:welfare_g_1}
   \end{eqnarray}
 \noindent Note also that if $b\in [\underline{b},b^\ast]$, it follows that:
       \begin{equation}
         c(t_r-t_l)\leq x(t_r)-x(t_l), \label{EQ:inequality}
       \end{equation}
        \noindent with strict inequality for $b\in [\underline{b},b^\ast)$. Finally, if $b>b^\ast$ the capacity $c$ is never exceeded so the welfare function (\ref{EQ:welfare_g_1}) can be simplified to:
\begin{equation}
T - \lambda g(b)(x(T)-x(0)).\label{EQ:not_exceeded}
\end{equation}

\medskip

\noindent \textbf{Proof of Proposition \ref{PROP:extreme_cases}.} Consider first part (i), and suppose that $b\in [\underline{b},b^\ast]$. By the above conclusions, it follows that $t_r>b>t_l>\underline{b}$ for all $b\in [\underline{b},b^\ast)$, and $t_l=t_r=b>\underline{b}$ for $b=b^\ast$. Furthermore, since $\lambda=0$ the welfare function (\ref{EQ:welfare_g_1}) can be written as:
       \begin{equation*}
         W( b)= T-(c(t_r-t_l)-(x(t_r)-x(t_l))).
       \end{equation*}
       \noindent From equation (\ref{EQ:inequality}), it follows that $W(b)<T$ for any $b\in [\underline{b},b^\ast)$ and $W( b)=T$ for $b=b^\ast$. If, on the other hand, $b>b^\ast$ and $\lambda=0$, the capacity is never exceeded so $W(b)=T$ by equation (\ref{EQ:not_exceeded}). This proves part (i) of the proposition.

To prove part (ii), note that since $\lambda=1$, the welfare function (\ref{EQ:welfare_g_1}) reduces to:
\begin{equation}
  W( b)= T - g(b)(x(T)-x(0)). \label{EQ:y_special}
\end{equation}
\noindent Since $g(\underline{b})\geq 1$ and $g^\prime(b)>0$ for all $b\geq \underline{b}$, and $x(T)-x(0)>0$, it follows that equation (\ref{EQ:y_special}) is maximized when $b$ is minimized. Hence, $b=\underline{b}$ maximizes equation (\ref{EQ:y_special}). \hfill $\square$

\medskip

\noindent \textbf{Proof of Proposition \ref{PROP:g_1}.} Consider first the case when $b\in [\underline{b},b^\ast]$. Because $g(\underline{b})=1$ and $g'(b)=0$ for all $b\geq \underline{b}$ by assumption, equation (\ref{EQ:welfare_g_1}) reduces to:
\begin{equation}
W(b)= T - \lambda (x(T)-x(0)) +(1-\lambda)(c(t_r-t_l)-(x(t_r)-x(t_l))).\label{EQ:welfare_g_0}
\end{equation}
\noindent Because $T-\lambda (x(T)-x(0))$ is a constant, equation (\ref{EQ:welfare_g_0}) is, by condition (\ref{EQ:inequality}), maximized when $b=b^\ast$, i.e., when $W( b^\ast)= T - \lambda (x(T)-x(0))$. To complete the proof, we need only to demonstrate that the welfare equals $T-\lambda (x(T)-x(0))$ for any $b>b^\ast$. But if $b>b^\ast$,  the capacity is never exceeded so the welfare function is given by equation (\ref{EQ:not_exceeded}) for $g(\underline{b})=1$ and $g'(b)=0$, i.e., $W(b)= T - \lambda (x(T)-x(0))$ for all $b>b^\ast$, which concludes the proof. \hfill $\square$

\medskip

\noindent \textbf{Proof of Proposition \ref{TH:main}.} Consider first the case when $b\in [\underline{b},b^\ast]$. Then the welfare function is given by equation (\ref{EQ:welfare_g_1}),
  \begin{equation*}
        W( b)= T - \lambda g(b)(x(T)-x(0)) +(1-\lambda)(c(t_r-t_l)-(x(t_r)-x(t_l))).
      \end{equation*}
\noindent For $b\geq b^\ast$, this equation reduces to $W( b)= T - \lambda g(b)(x(T)-x(0))$. Because $x(T)-x(0)$ is a constant, $g(b)\geq 1$ and $g^\prime(b)>0$ for all $b\geq \underline{b}$, it then follows that $W(b^\ast)>W(b)$ for any $b>b^\ast$. Hence, the welfare cannot be maximized for any $b>b^\ast$. \hfill $\square$

\bibliographystyle{apalike}
\bibliography{ref}

\end{document}